\documentclass[twoside]{article}

\usepackage{amsmath,amssymb,amsthm}

\newcommand{\N}{\mathbb{N}}
\newcommand{\C}{\mathbb{C}}
\newcommand{\R}{\mathbb{R}}
\newcommand{\B}{\mathcal{B}}
\newcommand{\M}{\mathcal{M}}
\newcommand{\K}{\mathcal{K}}
\newcommand{\ga}{\gamma}
\newcommand{\Ga}{\Gamma}
\newcommand{\la}{\lambda}
\newcommand{\La}{\Lambda}
\newcommand{\eps}{\varepsilon}
\newcommand{\1}{1\!\!1}
\newcommand{\Sect}{\mathrm{Sect}}
\newcommand{\sgn}{\mathrm{sgn}}
\newcommand{\const}{\mathrm{const}}
\renewcommand{\Re}{\mathrm{Re}}
\renewcommand{\Im}{\mathrm{Im}}
\newcommand{\LCM}{L_{\mathrm{CM}}}
\DeclareMathOperator*{\esssup}{ess\,sup}

\numberwithin{equation}{section}

\newtheorem{theorem}{Theorem}[section]
\newtheorem{lemma}[theorem]{Lemma}

\newtheorem{proposition}[theorem]{Proposition}
\newtheorem{remark}{Remark}[section]

\allowdisplaybreaks[3]

\title{INDIVIDUAL BASED MODEL WITH COMPETITION IN~SPATIAL~ECOLOGY\thanks{The financial support of DFG through the SFB 701
(Bielefeld University) and German-Ukrainian Project 436 UKR 113/80
and 436 UKR 113/94 is gratefully acknowledged. This work was
partially supported by FCT, POCI2010, FEDER. Yu.K. is very thankful
to R.~Law for fruitful and stimulating discussions.}}

\author{Dmitri Finkelshtein\thanks{Institute of Mathematics,
National Academy of Sciences of Ukraine, Kyiv, Ukraine ({\tt
fdl@imath.kiev.ua}).} \and Yuri Kondratiev\thanks{Fakult\"{a}t
f\"{u}r Mathematik, Universit\"{a}t Bielefeld, 33615 Bielefeld,
Germany ({\tt kondrat@math.uni-bielefeld.de}); Reading University.}
\and Oleksandr Kutoviy\thanks{Fakult\"{a}t f\"{u}r Mathematik,
Universit\"{a}t Bielefeld, 33615 Bielefeld, Germany ({\tt
kutoviy@math.uni-bielefeld.de}).}}

\begin{document}

\maketitle

\begin{abstract}
We analyze an interacting particle system with a Markov evolution of
birth-and-death type. We have shown that a local competition
mechanism (realized via a density dependent mortality) leads to a
globally regular behavior of the population in course of the
stochastic evolution.
\end{abstract}


{\small {\bf AMS subject classification.} 60J80; 60K35; 82C21; 82C22

\pagestyle{myheadings}
\thispagestyle{plain}
\markboth{D.~Finkelshtein, Yu.~Kondratiev, O.~Kutoviy}{IBM Model
with Competition in Spatial Ecology}

\section{Introduction}

Complex systems theory is a quickly growing interdisciplinary area with a very broad spectrum
of motivations and applications. Having in mind biological applications, S.~Levin (see \cite{Lev}) characterized
complex adaptive systems by such properties as
diversity and individuality of components,
localized interactions among components, and
the outcomes of interactions used for replication or
enhancement of components.
In the study of these systems,  proper language and techniques are delivered by the interacting particle models which form a rich and powerful direction in modern stochastic and infinite dimensional analysis.  Interacting particle systems have  a wide use as models in
condensed matter physics,
chemical kinetics,
population biology, ecology (individual based models),
sociology and economics (agent based models).

In this paper we  consider an individual based model (IBM) in
spatial ecology introduced by Bolker and Pacala \cite{BP97,BP99},
Dieckmann and Law \cite{DL00} (BDLP model). A population in this
model  is represented by a configuration of motionless organisms
(plants) located in an infinite habit (an Euclidean space in our
considerations). The habit is considered to be a continuous space as
opposed to discrete spatial lattices used in the most of
mathematical models of interacting particle systems. We need the
infinite habit  to avoid boundary effects in the population
evolution and the latter moment is quite similar to the necessity to
work in the thermodynamic limit for models of statistical physics.
Let us also mention a recent paper \cite{BiYo} in which a
modification of the BDLP model for the case of moving organisms
(e.g., branching diffusion of the plankton) was considered.

A general IBM in the plant ecology is a stochastic Markov process in
the configuration space with events comprising birth and death of
the configuration points, i.e., we are dealing with a
birth-and-death process in the continuum. In the particular case of
the BDLP model, each plant produces seeds independently of others
and then these seeds are distributed in the space accordingly to a
dispersion kernel $a^+$. This part of the process may be considered
as a kind of the spatial branching. In the same time, the model
includes also a mortality mechanism. The mortality intensity
consists of two parts. The first one corresponds to a constant
intrinsic mortality value $m>0$ s.t. any plant dies independently of
others after a random time (exponentially distributed with parameter
$m$). The second part in the mortality rate is density dependent.
The latter is expressed in terms of a competition kernel $a^-$ which
describes an additional mortality rate for any given point of the
configuration coming from the rest of the population, see Section 3
for the precise description of the model, in particular, (3.6). The
latter formula gives the heuristic form of the Markov generator in
the BDLP model.

Assuming the existence of the corresponding Markov process, we
derive in Section 5 an evolution equation for correlation functions
$k_{t}^{(n)}, n\geq 1,$ of the considered model. In
\cite{BP97,BP99}, \cite{DL00} this system was called  the system of
spatial moment equations for plant competition and, actually, this
system itself was taking as a definition of the dynamics in the BDLP
model. The mathematical structure of the correlation functions
evolution equation is close to other well-known hierarchical systems
in mathematical physics, e.g., BBGKY hierarchy for the Hamiltonian
dynamics (see, e.g. \cite{DSS}) or the diffusion hierarchy for the
gradient stochastic dynamic in the continuum (see e.g.
\cite{KoRRe}).  As in all hierarchical chains of equations, we can
not expect the explicit form of the solution, and even more, the
existence problem for these equations is a highly delicate question.

There is an approximative  approach to produce an information about
the behavior of the solutions to the hierarchical chains. This
approach is called the closure procedure and consists of the
following steps. The first step is to cut all correlation functions
of the higher orders and the second one is to subscribe the rest
correlation functions by the properly factorized correlation
functions of the lower orders. As result, one obtains a finite
system of non-linear equations instead of the original linear but
infinite system of a hierarchical type. This closure procedure is
essentially non-unique, see \cite{DL}.

The aim of this paper is to study the moment equations for the BDLP
model by methods of functional analysis and analysis on the
configuration spaces developed in  \cite{KoKu99}, \cite{KoKuKut},
\cite{KoKut} and already applied to the non-equilibrium
birth-and-death type continuous space stochastic dynamics in
\cite{KoKutMi}, \cite{KoKutZh}. We obtain some rigorous  results
concerning the existence and properties of the solution for
different classes of initial conditions.  One of the main question
we clarify in the paper concerns the role of the competition
mechanism in the regulation of the spatial structure of an evolving
population. More precisely, considering the model without
competition, i.e., the case $a^-\equiv 0$, we arrive in the
situation of the so-called continuous contact model \cite{FiKoSk},
\cite{KoKutPi07}, \cite{KS06}. In the ecological framework, this
model describes free growth of a plant population with the given
constant mortality. We note that (independently on the value of the
mortality $m>0$) the considered contact model exhibits very strong
clustering that is reflected in the bound (3.5) on the correlation
functions at any moment of time $t>0$. Note that this effect on the
level of the computer simulation was discovered already in
\cite{BiYo} and now it has the rigorous mathematical formulation and
clarification. A direct consequence of the competition in the model
is the suppression of such clustering. Namely, assuming the strong
enough competition and the big intrinsic mortality $m$, we prove the
sub-Poissonian bound for the solution to the moment equations
provided such bound was true for the initial state. Moreover, we
clarify specific influences of the constant and the density
dependent mortality intensities separately. More precisely, the big
enough intrinsic mortality $m$ gives a uniform in time bound for
each correlation function and the strong competition results ensure
the regular spatial distribution of the typical configuration for
any moment of time that is reflected in the sub-Poissonian bound.
Joint influence of the intrinsic mortality and the competition leads
to the existence of the unique invariant measure for our model which
is just Dirac measure concentrated on the empty configuration. The
latter means that the corresponding  stochastic evolution of the
population is asymptotically exhausting.

We would like to mention also the work \cite{FoMe} in which the BDLP
model was studied in the case of the bounded habit in the stochastic
analysis framework. The latter case differs essentially from the
model we consider in the present paper as well as main problems
studied in \cite{FoMe}, which are related to the scaling limits for
the considered processes.

\section{General facts and notations}

Let $\B({\R}^{d})$ be the family of all Borel sets in $\R^d$.
$\B_{b}({\R}^{d})$ denotes the system of all bounded sets in
$\B({\R}^{d})$.

The space of $n$-point configuration is
\[
\Ga _{0}^{(n)}=\Ga _{0,{\R}^{d}}^{(n)}:=\left\{ \left. \eta \subset
{\R}^{d}\right| \,|\eta |=n\right\} ,\quad n\in \N_0:=\N\cup \{0\},
\]
where $|A|$ denotes the cardinality of the set $A$. The space
$\Ga_{\La}^{(n)}:=\Ga _{0,\La }^{(n)}$ for $\La \in \B_b({\R}^{d})$
is defined analogously to the space $\Ga_{0}^{(n)}$. As a set, $\Ga
_{0}^{(n)}$ is equivalent to the symmetrization of
\[
\widetilde{({\R}^{d})^n} = \left\{ \left. (x_1,\ldots ,x_n)\in
({\R}^{d})^n\right| \,x_k\neq x_l\,\,\mathrm{if} \,\,k\neq l\right\}
,
\]
i.e. $\widetilde{({\R}^{d})^n}/S_{n}$, where $S_{n}$ is the
permutation group over $\{1,\ldots,n\}$. Hence one can introduce the corresponding topology and Borel $\sigma $- algebra, which we denote by $O(\Ga_{0}^{(n)})$ and $\B(\Ga_{0}^{(n)})$, respectively.
Also one can define a measure
$m^{(n)}$ as an image of the product of Lebesgue measures
$dm(x)=dx$ on $\bigl(\R^d, \B(\R^d)\bigr)$.

The space of finite configurations
\[
\Ga _{0}:=\bigsqcup_{n\in \N_0}\Ga _{0}^{(n)}
\]
is equipped with the topology which has structure of disjoint union. Therefore, one can define the corresponding Borel
$\sigma $-algebra $\B (\Ga _0)$.

A set $B\in \B (\Ga _0)$ is called
bounded if there exists $\La \in \B_b({\R}^{d})$ and $N\in \N$
such that $B\subset \bigsqcup_{n=0}^N\Ga _\La ^{(n)}$. The
Lebesgue---Poisson measure $\la_{z} $ on $\Ga_0$ is defined as
\[
\la _{z} :=\sum_{n=0}^\infty \frac {z^{n}}{n!}m ^{(n)}.
\]
Here $z>0$ is the so called activity parameter. The restriction of
$\la _{z} $ to $\Ga _\La $ will be also denoted by $\la _{z} $.

The configuration space
\[
\Ga :=\left\{ \left. \ga \subset {\R}^{d}\ \right| \; |\ga \cap \La
|<\infty, \text{ for all } \La \in \B_b({\R}^{d})\right\}
\]
is equipped with the vague topology. It is a Polish space (see e.g. \cite{KoKut}). The corresponding  Borel $\sigma $-algebra $ \B(\Ga )$ is defined as the smallest
$\sigma $-algebra for which all mappings $N_\La :\Ga \rightarrow
\N_0$, $N_\La (\ga ):=|\ga \cap \La |$ are measurable, i.e.,
\[
\B(\Ga )=\sigma \left(N_\La \left| \La \in
\B_b({\R}^{d})\right.\right ).
\]
One can also show that $\Ga $ is the projective limit of the spaces
$\{\Ga _\La \}_{\La \in \B_b({\R}^{d})}$ w.r.t. the projections $p_\La :\Ga \rightarrow \Ga _\La $, $p_\La (\ga ):=\ga _\La $, $\La \in \B_b({\R}^{d})$.

The Poisson measure $\pi
_{z} $ on $(\Ga ,\B(\Ga ))$ is given as the projective limit of the
family of measures $\{\pi _{z} ^\La \}_{\La \in \B_b({\R}^{d})}$,
where $\pi _{z} ^\La $ is the measure on $\Ga _\La $ defined by $\pi
_{z} ^\La :=e^{-z m (\La )}\la _{z}$.

We will use the following classes of functions:
$L_{\mathrm{ls}}^0(\Ga _0)$ is the set of all measurable functions
on $\Ga_0$ which have a local support, i.e. $G\in
L_{\mathrm{ls}}^0(\Ga _0)$ if there exists $\La \in \B_b({\R}^{d})$
such that $G\upharpoonright_{\Ga _0\setminus \Ga _\La }=0$;
$B_{\mathrm{bs}}(\Ga _0)$ is the set of bounded measurable functions
with bounded support, i.e. $G\upharpoonright_{\Ga _0\setminus B}=0$ for
some bounded $B\in \B(\Ga_0)$.

On $\Ga $ we consider the set of
cylinder functions $\mathcal{F}L^0(\Ga )$, i.e. the set of all
measurable functions $G$ on $\bigl(\Ga,\B(\Ga))\bigr)$ which are
measurable w.r.t. $\B_\La (\Ga )$ for some $\La \in \B_b({\R}^{d})$.
These functions are characterized by the following relation: $F(\ga
)=F\upharpoonright _{\Ga _\La }(\ga _\La )$.

The following mapping between functions on $\Ga _0$, e.g.
$L_{\mathrm{ls}}^0(\Ga _0)$, and functions on $\Ga $, e.g.
$\mathcal{F}L^{0}(\Ga )$, plays the key role in our further
considerations:
\begin{equation}
KG(\ga ):=\sum_{\eta \Subset \ga }G(\eta ), \quad \ga \in \Ga,
\label{KT3.15}
\end{equation}
where $G\in L_{\mathrm{ls}}^0(\Ga _0)$, see e.g.
\cite{KoKu99,Le75I,Le75II}. The summation in the latter expression
is taken over all finite subconfigurations of $\ga ,$ which is denoted by the symbol
$\eta \Subset \ga $. The mapping $K$ is linear, positivity
preserving, and invertible, with
\begin{equation}
K^{-1}F(\eta ):=\sum_{\xi \subset \eta }(-1)^{|\eta \setminus \xi
|}F(\xi ),\quad \eta \in \Ga _0.\label{k-1trans}
\end{equation}

Let $ \mathcal{M}_{\mathrm{fm}}^1(\Ga )$ be the set of all
probability measures $\mu $ on $\bigl( \Ga, \B(\Ga) \bigr)$ which
have finite local moments of all orders, i.e. $\int_\Ga |\ga _\La
|^n\mu (d\ga )<+\infty $ for all $\La \in \B_b(\R^{d})$ and $n\in
\N_0$. A measure $\rho $ on $\bigl( \Ga_0, \B(\Ga_0) \bigr)$ is
called locally finite iff $\rho (A)<\infty $ for all bounded sets
$A$ from $\B(\Ga _0)$. The set of such measures is denoted by
$\mathcal{M}_{\mathrm{lf}}(\Ga _0)$.

One can define a transform
$K^{*}:\mathcal{M}_{\mathrm{fm}}^1(\Ga )\rightarrow
\mathcal{M}_{\mathrm{lf}}(\Ga _0),$ which is dual to the
$K$-transform, i.e., for every $\mu \in
\mathcal{M}_{\mathrm{fm}}^1(\Ga )$, $G\in \B_{\mathrm{bs}}(\Ga _0)$
we have
\[
\int_\Ga KG(\ga )\mu (d\ga )=\int_{\Ga _0}G(\eta )\,(K^{*}\mu
)(d\eta ).
\]
The measure $\rho _\mu :=K^{*}\mu $ is called the correlation measure
of $\mu $.

As shown in \cite{KoKu99} for $\mu \in
\mathcal{M}_{\mathrm{fm}}^1(\Ga )$ and any $G\in L^1(\Ga _0,\rho
_\mu )$ the series \eqref{KT3.15} is $\mu $-a.s. absolutely
convergent. Furthermore, $KG\in L^1(\Ga ,\mu )$ and
\begin{equation}
\int_{\Ga _0}G(\eta )\,\rho _\mu (d\eta )=\int_\Ga (KG)(\ga )\,\mu
(d\ga ). \label{Ktransform}
\end{equation}

A measure $\mu \in \mathcal{M}_{\mathrm{fm} }^1(\Ga )$ is called
locally absolutely continuous w.r.t. $\pi _{z} $ iff $\mu_\La :=\mu
\circ p_\La ^{-1}$ is absolutely continuous with respect to $\pi
_{z} ^\La $ for all $\La \in \B_\La ({\R}^{d})$. In this case $\rho
_\mu :=K^{*}\mu $ is absolutely continuous w.r.t $\la _{z} $. We
denote
\[
k_{\mu}(\eta):=\frac{d\rho_{\mu}}{d\la_{z}}(\eta),\quad
\eta\in\Ga_{0}.
\]
The functions
\begin{equation}
k_{\mu}^{(n)}:(\R^{d})^{n}\longrightarrow\R_{+}
\end{equation}
\[ k_{\mu}^{(n)}(x_{1},\ldots,x_{n}):=\left\{\begin{array}{ll}
k_{\mu}(\{x_{1},\ldots,x_{n}\}), & \mbox{if $(x_{1},\ldots,x_{n})\in
\widetilde{(\R^{d})^{n}}$}\\ 0, & \mbox{otherwise}
\end{array}
\right.\] are the correlation functions well known in statistical
physics, see e.g \cite{R69}, \cite{ R70}.

We recall now the so-called Minlos lemma which plays very
important role in our calculations (cf., \cite{KoMiZh}).
\begin{lemma}\label{Minlos}
Let $n\in\N$, $n\geq 2$, and $z>0$ be given. Then
\begin{align*}
\int_{\Ga_{0}}\ldots\int_{\Ga_{0}}&G(\eta_{1}\cup\ldots\cup\eta_{n})H(\eta_{1},\ldots,\eta_{n})d\la_{z}(\eta_{1})\ldots
d\la_{z}(\eta_{n})\\
&=\int_{\Ga_{0}}G(\eta)\sum_{(\eta_{1},\ldots,\eta_{n})\in\mathcal{P}_{n}(\eta)}H(\eta_{1},\ldots,\eta_{n})d\la_{z}(\eta)
\end{align*}
for all measurable functions $G:\Ga_{0}\mapsto\R$ and
$H:\Ga_{0}\times\ldots\times\Ga_{0}\mapsto\R$ with respect to which
both sides of the equality make sense. Here $\mathcal{P}_{n}(\eta)$
denotes the set of all ordered partitions of $\eta$ in $n$ parts, which may
be empty.
\end{lemma}

\section{Description of the model}
In the present paper we study the special case of the general
birth-and-death processes in continuum. The spatial birth-and-death
processes describe evolution of configurations in $\R^d$, in which
points of configurations (particles, individuals, elements) randomly
appear (born) and disappear (die) in the space. Among all
birth-and-death processes we will distinguish those in which new
particles appear only from existing ones. These processes correspond
to the models of the spatial ecology.

The simplest example of such processes is the so-called ``free
growth'' dynamics. During this stochastic evolution the points of
configuration independently create new ones distributed in the space
according to a dispersion probability kernel $0\leq a^+\in
L^1(\R^d)$ which is an even function. Any existing point has an
infinite life time, i.\,e. they do not die. Heuristically, the
Markov pre-generator of this birth process has the following form:
\begin{equation*}
(L_{+}F)(\ga)=\varkappa^{+}\sum_{y\in\ga}\int_{\R^{d}}a^{+}(x-y)D_{x}^{+}F(\ga)dx,
\end{equation*}
where
\[
D_{x}^{+}F(\ga)=F(\ga\cup x)-F(\ga),
\]
and $\varkappa^{+}>0$ is some positive constant.

The existence of the process associated with $L_+$ can be shown
using the same technique as in \cite{FiKoSk}, \cite{KS06}. Let
$\mu_t$ be the corresponding evolution of measures in time on
$\mathcal{M}_{\mathrm{fm} }^1(\Ga )$. By $k_t^{(n)},\,n\geq 0$ we denote the
dynamics of the corresponding $n$-th order correlation functions
(provided they exist). Note, that each of such functions describes the
density of the system at the moment $t$.

Then, using \eqref{Ktransform}, for any continuous $\varphi$ on
$\R^d$ with bounded support, we obtain
\begin{align*}
\frac{d}{d t} \int_{\R^d} \varphi(x) k_t^{(1)}(x)dx &=\frac{d}{d t}
\int_\Ga \langle\varphi,\ga\rangle
d\mu_t(\ga)=\int_\Ga L_+\langle\varphi,\ga\rangle d\mu_t(\ga) \\
&=\varkappa^+ \int_\Ga \langle a^+*\varphi,\ga\rangle d\mu_t(\ga) =
\varkappa^+ \int_{\R^d} (a^+*\varphi)(x) k_t^{(1)}(x)dx \notag\\&=
\varkappa^+ \int_{\R^d} \varphi(x)(a^+*k_t^{(1)})(x)dx,\notag
\end{align*}
where $*$ denotes the classical convolution on ${\R^d}$. Hence,
$k_t^{(1)}$ grows exponentially in $t$. In particular, for the
translation invariant case one has $k_0^{(1)}(x)\equiv k_0^{(1)}>0$
and as a result
\begin{equation}\label{expgrowtn}
k_t^{(1)}=e^{\varkappa^+ t}k_0^{(1)}.
\end{equation}

One of the possibilities to prevent the density growth of the system
is to include the death mechanism. The simplest one is
described by the independent death rate (mortality) $m>0$. This means
that any element of a population has an independent exponentially
distributed with parameter $m$ random life time. The independent death
together with the independent creation of new particles by already
existing ones describe the so-called {\em contact model} in the continuum, see e.g. \cite{KS06}. The
pre-generator of such model is given by the following expression:
\begin{align*}
(\LCM F)(\ga)&= m \sum_{x\in\ga} D_{x}^{-}F(\ga)+(L_+F)(\ga)\\
&=m \sum_{x\in\ga}
D_{x}^{-}F(\ga)+\varkappa^{+}\sum_{y\in\ga}\int_{\R^{d}}a^{+}(x-y)D_{x}^{+}F(\ga)dx,
\end{align*}
where
\[
D_{x}^{-}F(\ga)=F(\ga\setminus x)-F(\ga).
\]

The Markov process associated with the generator $\LCM$ was constructed
in \cite{KS06}. This construction was generalized in \cite{FiKoSk}
for more general classes of functions $a^+$. Let us note, that the
contact model in the continuum may be used in the epidemiology to
model the infection spreading process. The values of this process
represent the states of the infected population. This is analog of
the contact process on a lattice. Of course, such interpretation is
not in the spatial ecology concept. On the other hand, contact process
is a spatial branching process with a given mortality rate.

The dynamics of correlation functions in the contact model was
considered in \cite{KoKutPi07}. Namely, taking $m=1$ for correctness, we have for
any $n \geq 1$, $t>0$ the correlation function of $n$-th order has
the following form
\begin{align}
k_{t}^{(n)}(x_{1},\ldots,x_{n})&= e^{n(\varkappa^+
-1)t}\left[\bigotimes_{i=1}^{n}e^{tL_{a^+}^{i}}\right]k_{0}^{(n)}(x_{1},\ldots,x_{n})
\label{evolt}
\\
&+\varkappa^+
\int_{0}^{t}e^{n(\varkappa^+-1)(t-s)}\left[\bigotimes_{i=1}^{n}e^{(t-s)L_{a^+}^{i}}\right]
\nonumber \\&\quad\times
\sum_{i=1}^{n}k_{s}^{(n-1)}(x_{1},\ldots,\check{x_{i}},\ldots,x_{n})\sum_{j:\,j\neq
i}a^+(x_{i}-x_{j})ds, \nonumber
\end{align}
where
\begin{align*}
&L_{a^+}^{i}k^{(n)}(x_{1},\ldots,x_{n})\\
=&\varkappa^+\int_{\R^{d}}a^+(x_{i}-y)
\left[k^{(n)}(x_{1},\ldots,x_{i-1},y,x_{i+1},
\ldots,x_{n})-k^{(n)}(x_{1},\ldots,x_{n})\right]dy
\end{align*}
and the symbol $\check{x_{i}}$ means that the $i$-th coordinate is
omitted. Note that $L_{a^+}^i$ is a Markov generator and the
corresponding semigroup (in $L^\infty$ space) preserves positivity.
It was also shown in \cite{KoKutPi07}, that if there exists a
constant $C>0$ (independent of $n$) such that for any $n\geq 0$
and $(x_{1},\ldots,x_{n})\in(\R^{d})^{n}$
\[
k_{0}^{(n)}(x_{1},\ldots,x_{n})\leq n!\,C^{n},
\]
then for any $t\geq 0$ and almost all (a.a.) $(x_{1},\ldots,x_{n})\in(\R^{d})^{n}$
(w.r.t. Lebesgue measure) the following estimate holds for all
$n\geq 0$
\begin{equation}
k_{t}^{(n)}(x_{1},\ldots,x_{n})\leq
\varkappa^+(t)^{n}(1+a_0)^{n}e^{n(\varkappa^+-1)t}(C+t)^{n}n!
\label{formula11}
\end{equation}
Here
\[
a_0=\|{a}\|_{\mathrm{L}^{\infty}(\R^{d})}, \qquad
\varkappa^+(t):=\max{\left[ 1,\,\varkappa^+,\,\varkappa^+
e^{-(\varkappa^+-1)t}\right]}.
\]

For the translation invariant case the value $\varkappa^+=1$ is
critical. Namely, from \eqref{evolt} we deduce that
\begin{equation}\label{CMgrowtn}
k_t^{(1)}=e^{(\varkappa^+-1) t}k_0^{(1)}.
\end{equation}
Therefore, for $\varkappa^+<1$ the density will exponentially
decrease to $0$ (as $t\rightarrow\infty$), for $\varkappa^+>1$ the
density will exponentially increase to $\infty$, and for
$\varkappa^+=1$ the density will be a constant.
One can easily see from the estimate \eqref{formula11} that, in the
case $\varkappa^+<1$, the correlation functions of all orders
decrease to $0$ as $t\rightarrow\infty$. On the other hand, for
fixed $t$, the estimate \eqref{formula11} implies factorial bound in
$n$ for $k_t^{(n)}$. As result, we may expect the clustering of our
system. To show clustering we start from the Poisson distribution of
particles and obtain an estimate from below for the time evolutions
of correlations between particles in a small region.

Hence, let $\varkappa^+<1$, $k_0^{(n)}=C^n$. Let $B$ is some bounded
domain of $\R^d$ such that
\[
\alpha:=\inf_{x,y\in B} a^+(x-y) >0.
\]
Let $\beta=\min\{\alpha\varkappa^+,C\}$. For any $\{x_1,x_2\}\subset
B$, formula \eqref{evolt} implies
\[
k_{t}^{(2)}(x_{1},x_{2})\geq 2C \varkappa^+\alpha
\int_{0}^{t}e^{2(\varkappa^+-1)(t-s)}ds\geq 2 \beta^2 t
e^{2(\varkappa^+-1)t}.
\]
We consider $t\geq 1$. One can prove by induction that for any
$\{x_1,\ldots,x_n\}\subset B$, $n\geq 2$
\begin{equation}\label{cluster}
k_t^{(n)}(x_1,\ldots,x_n)\geq \beta^n e^{n(\varkappa^+-1)t}n!
\end{equation}
Indeed, for $n=2$ this statement has been proved. Suppose that
\eqref{cluster} holds for $n-1$. Then, by \eqref{evolt}, one has
\begin{align*}
k_{t}^{(n)}(x_{1},\ldots,x_{n})&\geq \varkappa^+
\int_{0}^{t}e^{n(\varkappa^+-1)(t-s)} n \beta^{(n-1)}
e^{(n-1)(\varkappa^+-1)s}(n-1)! (n-1)\alpha ds\\
&\geq \beta^n n!
e^{n(\varkappa^+-1)t}\int_{0}^{t}e^{-(\varkappa^+-1)s}ds\geq \beta^n
e^{n(\varkappa^+-1)t} n! \,.
\end{align*}
As it was mentioned before, the later bound shows the clustering in
the contact model. All previous consideration may be extended for
the case $m\neq 1$: we should only replace $1$ by $m$ in the
previous calculations.

As a conclusion we have: the presence of mortality ($m>\varkappa^+$)
in the free growth model prevents the growth of density, i.\,e. the
correlation functions of all orders decay in time. But it doesn't
influence on the clustering in the system. One of the possibilities to
prevent such clustering is to consider the so-called density dependent
death rate. Namely, let us consider the following pre-generator:
\begin{align}
(LF)(\ga)&=
\sum_{x\in\ga}\left[m+\varkappa^{-}\sum_{y\in\ga\setminus
x}a^{-}(x-y)\right]D_{x}^{-}F(\ga)\label{BP-gen}\\&
+\varkappa^{+}\int_{\R^{d}}\sum_{y\in\ga}a^{+}(x-y)D_{x}^{+}F(\ga)dx.\nonumber
\end{align}
Here $0\leq a^{-}\in L^{1}(\R^{d})$ is an arbitrary, even function
such that
\[
\int_{\R^{d}}a^{-}(x)dx =1
\]
(in other words, $a^-$ is a probability density) and
$\varkappa^{-}>0$ is some positive constant.
It is easy to see that the operator $L$ is well-defined, for
example, on $\mathcal{F}L^{0}(\Ga)$.

The generator \eqref{BP-gen} describes the
Bolker---Dieckmann---Law---Pacala (BDLP) model, which was
introduced in \cite{BP97,BP99,DL00}. During the corresponding
stochastic evolution the birth of individuals occurs independently
and the death is ruled not only by the global regulation (mortality) but
also by the local regulation with the kernel $\varkappa^-a^-$. This
regulation may be described as a competition (e.g., for resources)
between individuals in the population.

The main result of this article is presented in
Section~\ref{s:evolCF}, Theorem \ref{5.1}. It may be informally stated in the following
way:

{\em If the mortality $m$ and the competition kernel
$\varkappa^-a^-$ are large enough, then the dynamics of correlation functions associated with
the pre-generator \eqref{BP-gen} preserves
(sub-)Poissonian bound for correlation functions for all times.}
\\
In particular, it prevents clustering in the model.

In the next sections we explain how to prove this fact. In the last
section of the present paper we discuss the necessity to consider "large enough" death.

\section{Semigroup for the symbol of the generator}

The problem of the construction of the corresponding process in $\Ga$
concerns the possibility to construct the semigroup associated with
$L$. This semigroup determines the solution to the Kolmogorov
equation, which formally (only in the sense of action of operator)
has the following form:
\begin{equation}
\frac{dF_t}{dt}=LF_t,\qquad F_t\bigm|_{t=0}=F_0.\label{Kolmogor}
\end{equation}

To show that $L$ is a generator of a semigroup in some reasonable
functional spaces on $\Ga$ seems to be a difficult problem. This
difficulty is hidden in the complex structure of the non-linear
infinite dimensional space $\Ga$.

In various applications the evolution of the corresponding
correlation functions (or measures) helps already to understand the
behavior of the process and gives candidates for invariant states.
The evolution of correlation functions of the process is related
heuristically to the evolution of states of our IPS. The latter
evolution is formally given as a solution to the dual Kolmogorov
equation (Fokker---Planck equation):
\begin{equation}
\frac{d\mu_t}{dt}=L^*\mu_t, \qquad \mu_t\bigm|_{t=0}=\mu_0,\label{FokkerPlanc}
\end{equation}
where $L^*$ is the adjoint operator to $L$ on
$\M^1_\mathrm{fm}(\Ga)$, provided, of course, that it exists.

In the recent paper \cite{KoKutMi}, the authors proposed the
analytic approach for the construction of a non-equilibrium process
on $\Ga$, which uses deeply the harmonic analysis technique. In the
present paper we follow the scheme proposed in \cite{KoKutMi} in
order to construct the evolution of correlation functions. The
existence problem for the evolution of states in
$\M^1_\mathrm{fm}(\Ga)$ and, as a result, of the corresponding
process on $\Ga$ is not realized in this paper. It seems to be a
very technical question and remains open.

Following the general scheme, first we should construct the
evolution of functions which corresponds to the \textit{symbol} ($K$-image)
$\hat{L}=K^{-1}LK$ of the operator $L$ in $L^1$-space on $\Ga_0$
w.r.t. the weighted Lebesgue---Poisson measure. This weight is
crucial for the corresponding evolution of correlation functions. It
determines the growth of correlation functions in time and space.
Below we start the detailed realization of the discussed scheme.

Let us set for $\eta\in\Ga_{0}$
\[
E^{a^{\#}}(\eta):=\sum_{x\in\eta}\sum_{y\in\eta\setminus
x}a^{\#}(x-y),
\]
where $a^{\#}$ denotes either $a^{-}$ or $a^{+}$.

\begin{proposition}\label{symbol}
The image of ${L}$ under the $K$-transform (or symbol of the
operator $L$) on functions $G\in B_{bs}(\Ga_{0})$ has the following
form
\begin{align*}
(\widehat{L}G)(\eta)&:=(K^{-1}L
KG)(\eta)\\&=-\left(m|\eta|+\varkappa^{-}E^{a^{-}}(\eta)\right)G(\eta)-\varkappa^{-}\sum_{x\in\eta}\sum_{y\in\eta\setminus
x}a^{-}(x-y)G(\eta\setminus y)\\
&+\varkappa^{+}\int_{\R^{d}}\sum_{y\in\eta}a^{+}(x-y)G((\eta\setminus
y)\cup
x)dx\\
&+\varkappa^{+}\int_{\R^{d}}\sum_{y\in\eta}a^{+}(x-y)G(\eta\cup
x)dx.
\end{align*}
\end{proposition}
For the proof see \cite{FiKoOl07}.

With the help of Proposition \ref{symbol}, we derive the evolution equation for \textit{quasi-observables} (functions on $\Ga_{0}$) corresponding to the Kolmogorov equation (\ref{Kolmogor}). It has the following form
\begin{equation}
\frac{dG_t}{dt}=\widehat{L}G_t,\qquad G_t\bigm|_{t=0}=G_0.\label{quasiKolmogor}
\end{equation}
Then in the way analogous to those in which the corresponding Fokker-Planck equation (\ref{FokkerPlanc}) was determined for (\ref{Kolmogor}) we get the evolution equation for the correlation functions corresponding to the equation (\ref{quasiKolmogor}):
\begin{equation}
\frac{dk_t}{dt}=\widehat{L}^{*}k_t,\qquad k_t\bigm|_{t=0}=k_0.\label{corrfunctiona}
\end{equation}
The precise form of the adjoint operator $\hat{L}^{*}$ 
will be given in Section \ref{s:evolCF}. It  is very important to emphasize that in the papers \cite{BP97,BP99} the equation (\ref{corrfunctiona}) was obtained from quit heuristic arguments and, moreover, it was considered as the definition for the evolution of the BDLP model.

Let $\la$ be the Lebesgue-Poisson measure on $\Ga_{0}$ with activity
parameter equal to 1.

For arbitrary and fixed $C>0$ we consider the operator $\widehat{L}$
as a pre-generator of a semigroup in the functional space
\begin{equation}
\mathcal{L}_{C}:=L^{1}(\Ga_{0}, C^{|\eta|}\la(d\eta)).\label{space1}
\end{equation}
In this section, symbol $\left\Vert\cdot\right\Vert_{C}$ stands for
the norm of the space (\ref{space1}).

For any $\omega>0$ we introduce the set $\mathcal{H}(\omega,0)$ of all
densely defined closed operators $T$ on $\mathcal{L}_{C},$ the
resolvent set $\rho(T)$ of which contains the sector
\[
\Sect\left(\frac{\pi}{2}+\omega\right):=\left\{\zeta\in\C\,\Bigm|
|\arg\, \zeta|<\frac{\pi}{2}+\omega\right\},\quad\omega>0
\]
and for any $\varepsilon >0$
\[
||(T-\zeta 1\!\!1)^{-1}||\leq \frac{M_{\varepsilon}}{|\zeta|},\quad
|\arg\,\zeta |\leq\frac{\pi}{2}+\omega-\varepsilon,
\]
where $M_{\varepsilon}$ does not depend on $\zeta$.

Let $\mathcal{H}(\omega,\theta)$, $\theta\in\R$ denotes the set of
all operators of the form $T=T_{0}+\theta$ with
$T_{0}\in\mathcal{H}(\omega,0)$.
\begin{remark}
It is well-known (see e.g., \cite{Kato}), that any
$T\in\mathcal{H}(\omega,\theta)$ is a generator of a semigroup
$U(t)$ which is holomorphic in the sector $|\arg\,t |<\omega$. The
function $U(t)$ is not necessarily uniformly bounded, but it is
quasi-bounded, i.e.
\[
||U(t)||\leq \const |e^{\theta t}|
\]
in any sector of the form $|\arg\,t|\leq \omega - \varepsilon$.
\end{remark}

\begin{proposition}\label{pr1} For any $C>0$, $m>0$, and $\varkappa^{-}>0$ the operator
\begin{gather*}
(L_{0}G)(\eta):=-\left(m|\eta|+\varkappa^{-}E^{a^{-}}(\eta)\right)G(\eta),\\
D(L_{0})= \left\{G\in\mathcal{L}_{C} \Bigm|
\left(m|\eta|+\varkappa^{-}E^{a^{-}}(\eta)\right)G(\eta)\in\mathcal{L}_{C}\right\}
\end{gather*}
is a generator of a contraction semigroup on $\mathcal{L}_{C}.$
Moreover, $L_{0}\in\mathcal{H}(\omega,0)$ for all $\omega\in
(0,\,\frac{\pi}{2})$.
\end{proposition}
\begin{proof} It is not difficult to show that $L_{0}$ is a densely
defined and closed operator in $\mathcal{L}_{C}.$

Let $0< \omega <\frac{\pi}{2}$ be arbitrary and fixed. Clear, that
for all $\zeta\in \Sect\left(\frac{\pi}{2}+\omega\right)$
\[
\bigl|m|\eta| + \varkappa^{-}E^{a^{-}}(\eta)+\zeta\bigr|>0,\quad
\eta\in\Ga_{0}.
\]
Therefore, for any $\zeta\in \Sect\left(\frac{\pi}{2}+\omega\right)$
the inverse operator $(L_{0}-\zeta 1\!\!1)^{-1}$, the action of
which is given by
\begin{equation}
[(L_{0}-\zeta
1\!\!1)^{-1}G](\eta)=-\frac{1}{m|\eta|+\varkappa^{-}E^{a^{-}}(\eta)+\zeta}\,G(\eta),
\label{necf}
\end{equation}
is well defined on the whole space $\mathcal{L}_{C}$. Moreover, it
is a bounded operator in this space and
\begin{equation}
||(L_{0}-\zeta 1\!\!1)^{-1}||\leq \left\{
\begin{array}{ll}
\frac{1}{|\zeta|}, & \mbox{if $\Re\,\zeta \geq 0$},\\
\\
\frac{M}{|\zeta|}, & \mbox{if $\Re\,\zeta<0$,}
\end{array}
\right.\label{bound}
\end{equation}
where the constant $M$ does not depend on $\zeta$.

The case $\Re\,\zeta \geq 0$ is a direct consequence of (\ref{necf})
and inequality
\[
m|\eta|+\varkappa^{-}E^{a^{-}}(\eta)+\Re\,\zeta \geq \Re\,\,\zeta
\geq 0.
\]
We prove now the bound (\ref{bound}) in the case $\Re\,\zeta < 0$.
Using (\ref{necf}), we have
\begin{align*}
||(L_{0}-\zeta 1\!\!1)^{-1}G||_{C}&=\left\Vert
\frac{1}{\bigl|m|\cdot|+\varkappa^{-}E^{a^{-}}(\cdot)+\zeta\bigr|}\,G(\cdot)\right\Vert_{C}=
\\
&=\frac{1}{|\zeta|}\left\Vert\frac{|\zeta|}{\bigl|
m|\cdot|+\varkappa^{-}E^{a^{-}}(\cdot)+\zeta\bigr|}G(\cdot)\right\Vert_{C}.
\end{align*}
Since $\zeta\in \Sect\left(\frac{\pi}{2}+\omega\right)$,
\[
|\Im\,\zeta|\geq
|\zeta|\left|\sin{\left(\frac{\pi}{2}+\omega\right)}\right|=|\zeta|\cos{\omega}.
\]
Hence,
\[
\frac{|\zeta|}{\bigl|m|\eta|+\varkappa^{-}E^{a^{-}}(\eta)+\zeta\bigr|}\leq
\frac{|\zeta|}{|\Im\,\zeta|}\leq\frac{1}{\cos{\omega}}=:M
\]
and (\ref{bound}) is fulfilled.

The rest of the statement of the lemma follows directly from the
theorem of Hille---Yosida (see e.g., \cite{Kato}).
\end{proof}

We define now
\[
(L_{1}G)(\eta):=\varkappa^{-}\sum_{x\in\eta}\sum_{y\in\eta\setminus
x}a^{-}(x-y)G(\eta\setminus y),\quad G\in D(L_{1}):=D(L_{0}).
\]
The lemma below implies that the operator $L_1$ is well-defined.
\begin{lemma}\label{lem1}
For any $\delta >0$ there exists $C_{0}:=C_{0}(\delta)>0$ such that
for all $C<C_{0}$ the following estimate holds
\begin{equation}
||L_{1}G||_{C}\leq a||L_{0}G||_{C}, \quad G\in D(L_{1}),
\label{pidporyad}
\end{equation}
with $a=a(C)<\delta$.
\end{lemma}
\begin{proof}
By modulus property
\begin{equation}
||L_{1}G||_{C}=
\int_{\Ga_{0}}\left|L_{1}G(\eta)\right|C^{|\eta|}\la(d\eta)\label{11}
\end{equation}
can be estimated by
\begin{equation}
\varkappa^{-}\int_{\Ga_{0}}\sum_{x\in\eta}\sum_{y\in\eta\setminus
x}a^{-}(x-y){|G(\eta\setminus y)|}C^{|\eta|}\la(d\eta).\label{1}
\end{equation}
By Minlos lemma, (\ref{1}) is equal to
\begin{gather*}
\varkappa^{-}\int_{\Ga_{0}}\int_{\R^{d}}\sum_{x\in\eta}a^{-}(x-y){|G(\eta)|}C^{|\eta|+1}dy\la(d\eta)=
\\
=\varkappa^{-}\int_{\Ga_{0}}{C|\eta|}|G(\eta)|C^{|\eta|}\la(d\eta)\leq
\frac{\varkappa^{-}}{m} C||L_{0}G||_{C}.
\end{gather*}
Therefore, \eqref{pidporyad} holds with
\[
a= \frac{\varkappa^{-}C}{m}.
\]
Clear, that taking
\[
C_{0} = \frac{\delta m}{\varkappa^{-}}
\]
we obtain that $a<\delta$ for $C<C_0$.
\end{proof}

We set now
\begin{align*}
(L_{2}G)(\eta)&:=(L_{2,\,\varkappa^{+}}G)(\eta)=\varkappa^{+}\int_{\R^{d}}\sum_{y\in\eta}a^{+}(x-y)G((\eta\setminus
y)\cup x)dx,\\ G\in D(L_{2})&:=D(L_{0}).
\end{align*}
The operator
$\bigl(L_2,D(L_2)\bigr)$ is well defined due to the lemma below:
\begin{lemma}\label{lem2}
For any $\delta >0$ there exists $\varkappa^{+}_{0}:=\varkappa^{+}_{0}(\delta)>0$
such that for all $\varkappa^{+}<\varkappa^{+}_{0}$ the following estimate holds
\begin{equation}
||L_{2}G||_{C}\leq a||L_{0}G||_{C}, \quad G\in D(L_{2}),
\label{pidporyad2}
\end{equation}
with $a=a(\varkappa^{+})<\delta$.
\end{lemma}
\begin{proof} Analogously to the previous lemma we estimate
\begin{equation} ||L_{2}G||_{C}=
\int_{\Ga_{0}}\left|L_{2}G(\eta)\right|C^{|\eta|}\la(d\eta)\label{12}
\end{equation}
by
\begin{equation}
\varkappa^{+}\int_{\Ga_{0}}\int_{\R^{d}}\sum_{y\in\eta}a^{+}(x-y){|G((\eta\setminus
y)\cup x)|}C^{|\eta|}dx\la(d\eta).\label{1a}
\end{equation}
By Minlos lemma, (\ref{1a}) is equal to
\[
\varkappa^{+}\int_{\Ga_{0}}\sum_{x\in\eta}\int_{\R^{d}}a^{+}(x-y){|G(\eta)|}C^{|\eta|}
dy\la(d\eta)\leq \frac{\varkappa^{+}}{m} ||L_0 G||_{C}.
\]
Taking $\varkappa^{+}_{0}=\delta m$ we prove the lemma.
\end{proof}

The operator defined as:
\begin{equation}
(NG)(\eta)=|\eta|G(\eta), \quad \quad G\in D(L_{0})
\end{equation}
is called the number operator.
\begin{remark}\label{rem1-2}
We proved, in particular, that for $G\in D(L_{0})=D(L_{1})=D(L_{2})$
\begin{align*}
||L_{1}G||_{C}&\leq\varkappa^{-}C||NG||_{C},\\
||L_{2}G||_{C}&\leq\varkappa^{+}||NG||_{C}.
\end{align*}
\end{remark}

Finally, we consider the last part of the operator $\widehat{L}$:
\[
(L_{3}G)(\eta):=\varkappa^{+}\int_{\R^{d}}\sum_{y\in\eta}a^{+}(x-y)G(\eta\cup
x)dx,\quad D(L_{3}):=D(L_{0}).
\]
\begin{lemma}\label{lem3}
For any $\delta >0$ and any $\varkappa^{+}>0$, $C>0$ such that
\begin{equation}\label{cond0}
\varkappa^{+}E^{a^{+}}(\eta)< \delta
C\left(\varkappa^{-}E^{a^{-}}(\eta)+m|\eta|\right)
\end{equation}
the following estimate holds
\begin{equation}
||L_{3}G||_{C}\leq a||L_{0}G||_{C}, \quad G\in D(L_{3}),
\label{pidporyad3}
\end{equation}
with $a=a(\varkappa^{+},\, C)<\delta$.
\end{lemma}
\begin{proof}
Using the same tricks as in the two previous lemmas we have
\begin{align}
||L_{3}G||_{C}&=
\int_{\Ga_{0}}\left|L_{3}G(\eta)\right|C^{|\eta|}\la(d\eta)\notag
\\ &\leq
\varkappa^{+}\int_{\Ga_{0}}\int_{\R^{d}}\sum_{y\in\eta}a^{+}(x-y){|G(\eta\cup
x)|}C^{|\eta|}dx\la(d\eta).\label{1b}
\end{align}
By Minlos lemma, (\ref{1b}) is equal to
\[
\frac{\varkappa^{+}}{C}\int_{\Ga_{0}}{E^{a^{+}}(\eta)|G(\eta)|}C^{|\eta|}
\la(d\eta).
\]
The assertion of the lemma is now trivial.
\end{proof}

\begin{theorem} \label{mainass} Assume that the functions $a^{-},\,a^{+}$ and the constants $\varkappa^{-}, \,\varkappa^{+}>0$, $m> 0$ and $C>0$ satisfy
\begin{equation}
C \varkappa^{-}a^{-}\geq 2\varkappa^{+}a^{+},\label{kappac}
\end{equation}
$$
m > 2\left(\varkappa^{-}C+\varkappa^{+}\right).
 $$
Then, the operator
$\widehat{L}$ is a generator of a holomorphic semigroup
$\hat{U}_{t},\, {t\geq 0}$ in $\mathcal{L}_{C}$.
\end{theorem}
\begin{proof}
The statement of the theorem follows directly from
Remark~\ref{rem1-2}, Lemma~\ref{lem3} and the theorem about
the perturbation of holomorphic semigroup (see, e.g. \cite{Kato}). For
the reader's convenience, below we give its formulation:

{\it For any $T\in\mathcal{H}(\omega,\,\theta)$ and for any
$\varepsilon > 0$ there exists positive constants $\alpha$, $\delta$
such that if the operator $A$ satisfies
\[
||Au||\leq a||Tu||+b||u||, \quad u\in D(T)\subset D(A),
\]
with $a<\delta$, $b<\delta$, then
$T+A\in\mathcal{H}(\omega-\varepsilon,\,\alpha)$.

In particular, if $\theta = 0$ and $b=0$, then $T+A\in
\mathcal{H}(\omega -\varepsilon,\,0)$}

Following the proof of this theorem (see, e.g. \cite{Kato}) and taking into account the fact that
$L_{0}\in\mathcal{H}(\omega,\,0)$ for any $\omega\in
(0,\,\frac{\pi}{2})$, one can conclude in our case that $\delta$ can be chosen equal to $\frac{1}{2}$. This is exactly the reason of
appearing multiplicand $2$  at the l.h.s. of \eqref{kappac}.
\end{proof}

\section{Evolution of correlation functions}\label{s:evolCF}
Let us consider the evolution equation (\ref{corrfunctiona}), which corresponds to the operator $\hat{L}^{*}$
\[
\frac{dk_t}{dt}=\hat{L}^{*}k_t,\qquad k_t\bigm|_{t=0}=k_0.
\]
Using the general scheme, proposed in \cite{FiKoOl07} we find the precise form of $\hat{L}^{*}$:
\begin{align*}
\hat{L}^{*}k(\eta)&=-\left(m|\eta|+\varkappa^{-}E^{a^{-}}(\eta)\right)k(\eta)
+\varkappa^{+}\sum_{x\in\eta}\sum_{y\in\eta\setminus
x}a^{+}(x-y)k(\eta\setminus x)\\
&+\varkappa^{+}\int_{\R^{d}}\sum_{y\in\eta}a^{+}(x-y)k((\eta\setminus
y)\cup x)dx\\
&-\varkappa^{-}\int_{\R^{d}}\sum_{y\in\eta}a^{-}(x-y)k(\eta\cup
x)dx.
\end{align*}
The main questions which we would like to study now are the existence and properties of the solution to the hierarchical  system of equations (\ref{corrfunctiona}). The answers to these questions are given in the following theorem
\begin{theorem}\label{5.1}
Suppose that all assumptions of Theorem \ref{mainass} are fulfilled. Then for any initial function $k_{0}$ from the class
\[
\K_{C}:=\left\{k:\Ga_{0}\rightarrow\R\,\left |\right.\,k\cdot
C^{-|\eta|}\in L^{\infty}(\Ga_{0},\la)\right\}
\]
the corresponding solution $k_{t}$ to (\ref{corrfunctiona}) exists and will be again the function from $\K_{C}$ for any moment of time $t\geq 0$.
\end{theorem}

 {\it Proof.} Following the scheme proposed in \cite{KoKutMi}, we construct the
corresponding evolution of the locally finite measures on $\Ga_{0}$. In
order to realize this construction we consider the dual space
$\K_{C}$
to the Banach space $\mathcal{L}_{C}$. The duality is given by the
following expression
\begin{equation}
\left\langle\!\left\langle G,\,k\right\rangle\!\right\rangle :=
\int_{\Ga_{0}}G\cdot k\, d\la,\quad G\in\mathcal{L}_{C} .
\label{duality}
\end{equation}
It is clear that $\K_{C}$ is the Banach space with the norm
\[
||k||:=||C^{-|\cdot|}k(\cdot)||_{L^{\infty}(\Ga_{0},\la)}.
\]
Note also, that $k\cdot C^{-|\cdot|}\in L^{\infty}(\Ga_{0},\la)$
means that the function $k$ satisfies the bound
\[
|k(\eta)|\leq \const C^{|\eta|} \quad \text{for } \la\text{-a.a. }
\eta\in\Ga_0.
\]
The evolution on $\K_{C}$, which corresponds to $\hat{U}_{t},\,
_{t\geq 0}$ constructed in Theorem \ref{mainass}, may be determined in the following way:
\[
\left\langle\!\left\langle G,\,k_{t}\right\rangle\!\right\rangle
:=\left\langle\!\!\left\langle
\hat{U}_{t}G,\,k_{0}\right\rangle\!\!\right\rangle.
\]
We denote
\[
\hat{U}_{t}^{*}k_{0}:=k_{t}.
\]
Using the same arguments as in \cite{KoKutMi}, it becomes clear that $k_{t}=\hat{U}_{t}^{*}k_{0}$ is the solution to (\ref{corrfunctiona}) in the Banach space $\K_{C}$.\hfill $\square$


It is important to emphasize that in the case of $a^{-}\equiv 0$ and
$\varkappa^{+}< m$
\[
k_{t}^{(n)}\rightarrow 0, \quad t\rightarrow
0,\quad \mathrm{for}\quad\mathrm{any}\quad n\geq1,
\]
see e.\,g. \cite{KoKutPi07}. Therefore, we may expect that
the correlation functions of our model satisfy this property as well.

\section{Stationary equation for the system of correlation functions}

Let us consider for any $\alpha\in\R$ the following Banach subspace
of $\K_C$:
\[
\K_C^\alpha:=\bigl\{ k\in\K_C \bigm| k^{(0)}(\emptyset)=\alpha
\bigr\}.
\]

In this section we study the existence problem for the solutions to the stationary equation
\begin{equation}
\hat{L}^{*}k=0\label{station}
\end{equation}
in $\K_C^1$. The main result is formulated in the following way:
\begin{theorem} Suppose that
\begin{equation}
\frac{C\varkappa^{-}}{m} +\frac{\varkappa^{+}}{m} +\frac{1}{C}<1\label{constantC}
\end{equation}
and
\[
\varkappa^{-} a^{-}\geq\varkappa^{+}a^{+}
\]
then the solution $k=(k^{(n)})_{n\geq 0}$ to (\ref{station}) is
unique in $\K_C^1$ and such that
\[
k^{(n)}=0,\quad\quad n\geq 1.
\]
\end{theorem}
\begin{proof}
Let
\[
\left( \hat{L}^{\star }k\right) \left( \eta \right) =0.
\]
The latter means that
\begin{eqnarray*}
&&\left( m\left\vert \eta \right\vert +\varkappa^{-}E^{a^{-}}\left(
\eta \right) \right) k\left( \eta \right)=-\varkappa^{-}\sum_{x\in
\eta }\int_{\R^{d}}k\left( y\cup \eta \right)
a^{-}\left( x-y\right) dy+ \\
&&+\varkappa^{+}\sum_{x\in\eta}\sum_{y\in\eta\setminus x}a^{+}(x-y)k(\eta\setminus x)+\varkappa^{+}\int_{\R^{d}}\sum_{y\in\eta}a^{+}(x-y)k((\eta\setminus y)\cup
x)dx.
\end{eqnarray*}
The last relation holds for any $k\in\K_C^1$ at the point
$\eta=\emptyset$. Hence, one can consider it on $\K_C^0$.

Let us denote for $\eta \neq \emptyset $
\begin{eqnarray*}
\left( Sk\right) \left( \eta \right)
&=&-\frac{\varkappa^{-}}{m\left\vert \eta \right\vert
+\varkappa^{-}E^{a^{-}}\left( \eta \right) }\sum_{x\in \eta
}\int_{\R^{d}}k\left( y\cup \eta \right) a^{-}\left( x-y\right) dy +\\
&&+\frac{\varkappa^{+}}{m\left\vert \eta \right\vert +\varkappa^{-}E^{a^{-}}\left( \eta \right) }
\sum_{x\in\eta}\sum_{y\in\eta\setminus x}a^{+}(x-y)k(\eta\setminus x)+\\
&&+\frac{\varkappa^{+}}{m\left\vert \eta \right\vert +\varkappa^{-}E^{a^{-}}\left( \eta \right) }
\int_{\R^{d}}\sum_{y\in\eta}a^{+}(x-y)k((\eta\setminus y)\cup
x)dx\\
\end{eqnarray*}
and
\[
\left( Sk\right) \left( \emptyset \right) =0.
\]
Let
\[
\left\Vert k\right\Vert _{C}=\esssup_{\eta \in \Ga
_{0}}\frac{\left\vert k\left( \eta \right) \right\vert
}{C^{\left\vert \eta \right\vert }},
\]
then
\begin{eqnarray*}
&&\left\Vert Sk\right\Vert _{C} \\
&\leq &\left\Vert k\right\Vert _{C}\esssup_{\eta \in \Ga
_{0}\setminus \left\{ \emptyset \right\}
}\frac{\varkappa^{-}C}{m\left\vert \eta \right\vert
+\varkappa^{-}E^{a^{-}}\left( \eta \right) }\sum_{x\in \eta
}\int_{\R^{d}}a^{-}\left( x-y\right) dy \\
&&+\frac{\left\Vert k\right\Vert _{C}}{C}\,\esssup_{\eta \in \Ga
_{0}\setminus \left\{ \emptyset \right\}
}\frac{\varkappa^{+}}{m\left\vert \eta \right\vert
+\varkappa^{-}E^{a^{-}}\left( \eta \right) }\sum_{x\in\eta}\sum_{y\in\eta\setminus x}a^{+}(x-y)\\
&&+{\left\Vert k\right\Vert _{C}}{}\,\esssup_{\eta \in \Ga
_{0}\setminus \left\{ \emptyset \right\}
}\frac{\varkappa^{+}}{m\left\vert \eta \right\vert
+\varkappa^{-}E^{a^{-}}\left( \eta \right) }
\int_{\R^{d}}\sum_{y\in\eta}a^{+}(x-y)dx\\
&\leq &\left\Vert k\right\Vert _{C}\frac{C\varkappa^{-} }{m
}+\left\Vert k\right\Vert _{C}\frac{\varkappa^{+} }{m} +\left\Vert k\right\Vert _{C}
\frac{1}{C}=\left\Vert k\right\Vert _{C}\left( \frac{C\varkappa^{-}}{m} +\frac{\varkappa^{+}}{m} +\frac{1}{C}\right),
\end{eqnarray*}
if
\[
\varkappa^{+}E^{a^{+}}\left( \eta \right) \leq\varkappa^{-} E^{a^{-}}\left( \eta \right) +m\left\vert
\eta \right\vert.
\]
As result,
\[
\left\Vert S\right\Vert \leq \frac{1}{m}C\varkappa^{-} +
\frac{1}{m}\varkappa^{+} +\frac{1}{C}<1.
\]
The assertion of the theorem is now obvious.
\end{proof}
\begin{remark}
For any $C>1$ one may chose $\varkappa^{-}>0$ and $m>0$ such that (\ref{constantC}) is satisfied. The latter means, that, asymptotically, our system exhausted to the system with the stationary state $\delta_{\emptyset}(d\ga)$ (the Dirac measure concentrated on the empty configuration $\emptyset$). In other words, the population evolving due to the BDLP dynamics is asymptotically degenerated.
\end{remark}

\section{Further developments}

In Theorem \ref{5.1} we have shown that functions $k_t$ is
bounded by $C^n$ for all $t>0$, provided that $k_0$ satisfies initially the bound of the same type.
Using approximation arguments (see e.g. \cite{KoKutMi}, \cite{KoKutZh}) one may prove that the
corresponding time evolution of the correlation function will be
also correlation function for some probability measure on $\Ga$. We
suppose to discuss this problem as well as other probabilistic aspects of the BDLP model in a forthcoming paper.
The main aim of the present paper is to analyze evolution of correlation functions. Namely, we have shown that dynamics
of correlation functions stays in the space $\K_C$. This property
seems to be very strong. To show that system of correlation
functions evolving in time stays in the same space is already
difficult even for the contact model. Namely, \eqref{formula11} implies that the evolution of correlation
functions at some moment of time $t$ may leave the space
\[
\left\{k:\Ga_{0}\rightarrow\R\,\left |\right.\,k\cdot
C^{-|\eta|}\cdot|\eta|!\in L^{\infty}(\Ga_{0},\la)\right\}.
\]
The reason is that $C$ may depend on $t$, which is true at least for the case
$\varkappa^+\geq 1$ ($m=1$ at the moment). Hence, we may expect that
the dynamics of correlation functions for the contact process lives
in some bigger space. Of course, this is possible only for
$\varkappa^+\leq 1$ since for $\varkappa^+>1$ density tends to
infinity. Hence, let us consider the case $\varkappa^+=1$. One
candidate for such bigger space is
\[
\mathcal{R}_C:=\left\{k:\Ga_{0}\rightarrow\R\,\left |\right.\,k\cdot
C^{-|\eta|}\cdot (|\eta|!)^2 \in L^{\infty}(\Ga_{0},\la)\right\}.
\]
Note, that the invariant measure of the contact process
belongs to this space
(see \cite[Theorem 4.2]{KoKutPi07}), provided that $d\geq 3$, $a^+$
has finite second moment w.r.t. the Lebesgue measure and the Fourier
transform of $a^+$ is integrable on $\R^d$. Below we show that the
evolution of correlation functions at any moment of time $t$ is a
function from $\mathcal{R}_C$.

Indeed, let $\varkappa^+=1$ and suppose that there exists $C>0$ such
that for any $n\geq1$ and for any $x_1,\ldots,x_n\in\R^d$
\[
k_0^{(n)}(x_1,\ldots,x_n)\leq\frac{1}{2} C^n (n!)^2.
\]
Then, it is clear that $k_0\in\mathcal{R}_C$. Now, suppose that
$k_t^{(n-1)}\leq C^{n-1} ((n-1)!)^2$. We prove the corresponding
inequality for $k_t^{(n)}$ using the mathematical induction. By
\eqref{formula11} we have for any $x_1,\ldots,x_n\in\R^d$
\begin{align}
&k_t^{(n)}(x_1,\ldots,x_n)\label{e:ddd}\\&\leq \frac{1}{2} C^n
(n!)^2\nonumber
\\
&\quad+\int_{0}^{t}
\left[\bigotimes_{i=1}^{n}e^{(t-s)L_{a^+}^{i}}\right]
\sum_{i=1}^{n}C^{n-1} ((n-1)!)^2\sum_{j:\,j\neq
i}a^+(x_{i}-x_{j})ds\nonumber\\
&=\frac{1}{2} C^n (n!)^2 + C^{n-1} ((n-1)!)^2
\sum_{i=1}^{n}\sum_{j:\,j\neq i}\int_0^t\left(e^{2(t-s)L_{a^+}}
a^+\right)(x_{i}-x_{j})ds,\nonumber
\end{align}
where for $f\in L^1(\R^d)$
\[
L_{a^+}f (x) = \int_{\R^d} a(x-y) [f(y)-f(x)]dx,\quad x\in\R^d.
\]
For the bound above we have used the fact, that for any $1\leq i\neq
j\leq n$
\[
L_{a^+}^{i} a^+(x_i-x_j)=L_{a^+}^{j} a^+(x_i-x_j) = (L_{a^+}
a^+)(x_i-x_j),\quad x_i, x_j\in\R^d.
\]
This relation can be easily checked by simple computations.

Note, that $L_{a^+}$ is a generator of the Markov semigroup which
preserves positivity in $L^1(\R^d)$. Hence,
\[
g_t(x):=\int_0^t\left(e^{2(t-s)L_{a^+}} a^+\right)(x)ds\geq 0,\quad
x\in\R^d, t\geq 0,
\]
and $g_t\in L^1(\R^d)$. Then we have
\[
g_t(x)=|g_t(x)|= \frac{1}{(2\pi)^d}\left|\int_{\R^d} e^{ipx}
\widehat{g}_t(p) dp\right| \leq
\frac{1}{(2\pi)^d}\int_{\R^d}\int_0^t
e^{2(t-s)(\widehat{a}^+(p)-1)}|\widehat{a}^+(p)|dsdp,
\]
where symbol $\widehat{f}$ denotes the Fourier transform of the function
$f\in L^1(\R^d)$. Therefore,
\[
g_t(x)\leq \frac{1}{(2\pi)^d}\int_{\R^d}
\frac{1-e^{2t(\widehat{a}^+(p)-1)}}{2(1-\widehat{a}^+(p))}|\widehat{a}^+(p)|dp
\leq \frac{1}{2(2\pi)^d}\int_{\R^d}
\frac{|\widehat{a}^+(p)|}{1-\widehat{a}^+(p)}dp.
\]
It was shown in \cite{KoKutPi07} that under the conditions posed on
function $a^+$ for the case of invariant measure
\[
D:=\int_{\R^d}
\frac{|\widehat{a}^+(p)|}{1-\widehat{a}^+(p)}dp<\infty.
\]
Finally, if additionally
\[
C\geq \frac{D}{(2\pi)^d},
\]
then we obtain from \eqref{e:ddd}
\[
k_t^{(n)}(x_1,\ldots,x_n)\leq \frac{1}{2} C^n (n!)^2 + \frac{1}{2}
C^{n-1} ((n-1)!)^2 n(n-1)\frac{D}{(2\pi)^d} \leq C^n (n!)^2.
\]
As result, $k_t\in\mathcal{R}_C$ for all $t\geq 0$.

Therefore, the dynamics of correlation functions for the contact
model stays in $\mathcal{R}_C$, hence, this dynamics is really very
clustering for $\varkappa^+=m=1$. As before, we may extend our
consideration on the case $m\neq 1$.

Summarizing previous results in this section we claim that the presence
of the big mortality and the big competition kernel prevents clustering
in the system making it sub-Poissonian distributed. But, is it
really necessary to add ``big'' mortality and competition kernel?
Below we discuss this problem.

If we want to study the quasibounded semigroup with the generator
$\widehat{L}$ on $\mathcal{L}_{C}$ for some $C>0$ then, naturally,
this generator should be an accretive operator in $\mathcal{L}_{C}$.
Hence, for some $b\geq 0$ the following bound should be true
\[
\int_{\Gamma _{0}}\sgn\left( G\left( \eta \right) \right) \cdot
\left( \bigl( \hat{L}-b\1 \bigr) G\right) \left( \eta \right)
d\lambda _{C}\left( \eta \right) \leq 0, \quad \forall\; G\in
D(\widehat{L}),
\]
since
\[
C^{|\eta|}d\lambda (\eta) =d\lambda _{C}(\eta).
\]
Let us define the ``diagonal'' part of the operator $\widehat{L}$:
\[
\bigl( \hat{L}_{diag}G\bigr) \left( \eta \right) :=-m\vert\eta\vert
G(\eta)-\varkappa ^{-}E^{a^{-}}(\eta )G(\eta )+\varkappa
^{+}\int_{\mathbb{R}^{d}}\sum_{y\in \eta }a^{+}(x-y)G((\eta
\setminus y)\cup x)dx
\]
and consider for some $n\geq 1$
\[
G=\left( 0,0,G^{\left( n\right) },0,0\right), \quad G^{(n)}\in
L^1((\R^d)^n).
\]
Then
\[
(\widehat{L}G)(\eta )=\left\{
\begin{array}{ll}
\varkappa ^{+}\int\limits_{\mathbb{R}^{d}}\sum\limits_{y\in \eta
}a^{+}(x-y)G^{\left(
n\right) }(\eta \cup x)dx,&\left\vert \eta \right\vert =n-1 \\
-\varkappa ^{-}\sum\limits_{x\in \eta }\sum\limits_{y\in \eta
\setminus x}a^{-}(x-y)G^{\left( n\right) }(\eta \setminus
y),&\left\vert \eta
\right\vert =n+1 \\
\left( \hat{L}_{diag}G^{\left( n\right) }\right) \left( \eta \right)
,&\left\vert \eta \right\vert =n \\
0,&\text{ otherwise}
\end{array}
\right. .
\]
Note that $\sgn\left( G\left( \eta \right) \right)\equiv0$ if
$\vert\eta\vert\neq n$.

Therefore, for arbitrary $n\geq 1$
\begin{align*}
0\geq I_{n} &:=\int_{\Gamma _{0}}\sgn\left( G\left( \eta \right)
\right) \cdot \Bigl( \bigl( \hat{L}-b\1 \bigr) G\Bigr) \left( \eta
\right)
d\lambda _{C}\left( \eta \right)  \\
&=\int_{\Gamma _{0}^{(n)}}\sgn\left( G\left( \eta \right) \right)
\cdot \Bigl( \bigl( \hat{L}_{diag}-b\1 \bigr) G^{\left( n\right)
}\Bigr)\left(
\eta \right) d\lambda _{C}\left( \eta \right)  \\
&=\frac{C^{n}}{n!}\int_{\left( \mathbb{R}^{d}\right) ^{n}}\sgn\Bigl(
G^{\left( n\right) }\bigl( x^{( n) }\bigr) \Bigr) \Bigl( \bigl(
\hat{L}_{diag}-b\1 \bigr) G^{\left( n\right) }\Bigr) \bigl( x^{( n)
}\bigr) dx^{\left( n\right) }.
\end{align*}

Let us fix some $t>0$ and $\La\in\B_b(\R^d)$. Set for $n\geq 1$
\[
G^{\left( n\right) }\bigl( x^{( n) }\bigr)
=t^{n}\prod\limits_{k=1}^{n}\chi _{\Lambda }\left( x_{k}\right)
=t^n\1_{\Ga_\La^{(n)}}\bigl(\{x^{(n)}\}\bigr)\in L^1((\R^d)^n).
\]
Then, the equality
\[
\sgn\Bigl( G^{\left( n\right) }\bigl( x^{( n) }\bigr) \Bigr)
=\prod\limits_{k=1}^{n}\chi _{\Lambda }\left( x_{k}\right)
\]
implies
\begin{align*}
&0\geq \frac{n!}{t^{n}C^{n}}I_{n} \\
&=\int_{\Lambda ^{n}}\left( -mn \prod\limits_{k=1}^{n}\chi _{\Lambda
}\left( x_{k}\right)-\varkappa ^{-}E^{a^{-}} \bigl( x^{( n) }\bigr)
\prod\limits_{k=1}^{n}\chi _{\Lambda }\left( x_{k}\right) \right. \\
& \qquad+\left. \varkappa
^{+}\int_{\mathbb{R}^{d}}\sum_{j=1}^{n}a^{+}(x-x_{j})\prod
\limits_{k\neq j}\chi _{\Lambda }\left( x_{k}\right) \chi _{\Lambda
}\left( x\right) dx\right) dx^{\left( n\right) }-b\int_{\Lambda
^{n}}\prod\limits_{k=1}^{n}\chi _{\Lambda }\left( x_{k}\right)
dx^{\left(
n\right) } \\
&=-\varkappa ^{-}\int_{\Lambda ^{n}}E^{a^{-}}\bigl( x^{( n) }\bigr)
dx^{\left( n\right) }+\varkappa
^{+}\sum_{j=1}^{n}\prod\limits_{k\neq j}\int_{\Lambda
^{n-1}}dx_{k}\int_{\Lambda }\int_{\Lambda }a^{+}(x-x_{j})dxdx_{j}\\
& \quad -(b+mn)
\left\vert \Lambda \right\vert ^{n} \\
&=-\varkappa ^{-}\int_{\Lambda ^{n}}E^{a^{-}}\bigl( x^{( n)
}\bigr)dx^{\left( n\right) }+\varkappa ^{+}n\left\vert \Lambda
\right\vert ^{n-1}\int_{\Lambda }\int_{\Lambda
}a^{+}(x-y)dxdy-(b+mn)\left\vert \Lambda \right\vert ^{n}.
\end{align*}
We suppose, in fact, that for any $n\geq 1$
\[
I_{n}\leq 0.
\]
Since $E^{a^{-}}\left( \eta \right) =0$ for $\left\vert \eta
\right\vert \leq 1$ we get
\begin{align*}
0 &\geq \sum_{n=1}^{\infty }I_{n}=-m\sum_{n=1}^{\infty
}n\frac{t^{n}C^{n}}{n!} \left\vert \Lambda \right\vert
^{n}-\varkappa ^{-}\sum_{n=1}^{\infty }\frac{
t^{n}C^{n}}{n!}\int_{\Lambda ^{n}}E^{a^{-}}\bigl( x^{( n) }\bigr)
dx^{\left( n\right) }\\ & \quad +\varkappa ^{+}\sum_{n=1}^{\infty
}\frac{ t^{n}C^{n}}{n!}n\left\vert \Lambda \right\vert
^{n-1}\int_{\Lambda }\int_{ \Lambda
}a^{+}(x-y)dxdy-b\sum_{n=1}^{\infty }\frac{t^{n}C^{n}}{n!}
\left\vert \Lambda \right\vert ^{n} \\
&=-mtC\left\vert \Lambda \right\vert e^{Ct\left\vert \Lambda
\right\vert }-\varkappa ^{-}\int_{\Gamma _{\Lambda }}E^{a^{-}}\left(
\eta \right) d\lambda _{Ct}\left( \eta \right) +\varkappa
^{+}Cte^{Ct\left\vert \Lambda \right\vert }\int_{\Lambda
}\int_{\Lambda }a^{+}(x-y)dxdy\\&\quad -b\left(
e^{Ct\left\vert \Lambda \right\vert }-1\right)  \\
&=-mtC\left\vert \Lambda \right\vert e^{Ct\left\vert \Lambda
\right\vert }-\varkappa ^{-}C^{2}t^{2}\int_{\Gamma _{\Lambda
}}\int_{\Lambda }\int_{\Lambda }a^{-}\left( x-y\right) dxdyd\lambda
_{Ct}\left( \eta \right)
\\ &\quad +\varkappa ^{+}Cte^{Ct\left\vert \Lambda \right\vert }\int_{\Lambda }
\int_{\Lambda }a^{+}(x-y)dxdy-b\left( e^{Ct\left\vert \Lambda
\right\vert }-1\right)  \\ &=e^{Ct\left\vert \Lambda \right\vert
}\left[ Ct\left( \varkappa ^{+}\int_{\Lambda }\int_{\Lambda
}a^{+}(x-y)dxdy-\varkappa ^{-}Ct\int_{\Lambda }\int_{\Lambda
}a^{-}\left( x-y\right) dxdy-m\left\vert \Lambda \right\vert\right)
\right. \\ & \quad \left. -b\left( 1-e^{-Ct\left\vert \Lambda
\right\vert }\right) \vphantom{\int_{\Lambda }}\right] .
\end{align*}
Therefore, for any $t>0$ and any $\La\in\B_b(\R^d) $
\begin{align*}
0&\geq \varkappa ^{+}\int_{\Lambda }\int_{\Lambda
}a^{+}(x-y)dxdy-\varkappa ^{-}Ct\int_{\Lambda }\int_{\Lambda
}a^{-}\left( x-y\right) dxdy-m\left\vert \Lambda \right\vert \\
&\quad -b\frac{\left( 1-e^{-Ct\left\vert \Lambda \right\vert
}\right)}{Ct}=:B.
\end{align*}
Suppose that there exists $z>0$ such that
\[
a^{+}\left( x\right) \geq za^{-}\left( x\right) ,\quad x\in
\mathbb{R}^{d},
\]
then taking for some $\eps>0$
\[
t=\eps\frac{z\varkappa ^{+}}{\varkappa ^{-}C}>0
\]
we obtain
\begin{align*}
B&\geq (1-\eps)\varkappa ^{+}z \int_{\Lambda }\int_{\Lambda
}a^{-}(x-y)dxdy-m\left\vert \Lambda \right\vert\\&\quad
-\frac{b\varkappa ^{-}}{\eps z\varkappa ^{+}}\left( 1-\exp \left( -
\frac{z\varkappa ^{+}}{2\varkappa ^{-}}\left\vert \Lambda
\right\vert \right) \right)  \sim \Bigl((1-\eps)\varkappa ^{+}z
-m\Bigr) \vert\La\vert , \quad \Lambda \uparrow \mathbb{R}^{d},
\end{align*}
which contradicts to $B\leq 0$. As result, $m$ can not be
arbitrary small.

\end{document}